\documentclass[aps,prb,twocolumn,superscriptaddress,10pt]{revtex4}
\usepackage{amsfonts}
\usepackage{amsmath}
\usepackage{amssymb}
\usepackage{graphicx}
\usepackage{dcolumn}
\usepackage{mciteplus}
\usepackage{euscript}
\usepackage{multirow}

\providecommand{\U}[1]{\protect\rule{.1in}{.1in}}
\def\efermi{$E_{\rm F}$}
\def\mave{${M}$}
\def\mscell{$M$}
\def\mspin{$m_{i}$}
\def\msave{${m}$}
\def\morb{$l$}
\def\desoi{$\Delta_{i}$}
\def\deso{$\Delta$}
\def\deg{$^{\circ}\mathrm{C}$}

\def\jij{{$J_{ij}$}}

\def\fen{Fe$_{16}$N$_2$}
\def\afen{$\alpha''$-Fe$_{16}$N$_{2}$}
\def\mub{$\mu_B$}

\def\fee{${4e}$}
\def\feh{${8h}$}
\def\fed{${4d}$}

\def\tc{{$T_\text{C}$}}
\def\gw{\emph{GW}}
\def\qsgw{\mbox{QS$GW$}}
\def\ldau{LDA+$U$}
\def\etal{$\textit{et al.}$}

\newcommand{\rfig}[1]{Fig.~\ref{#1}}

\begin{document}

\title{Effects of alloying and strain on the magnetic properties of Fe$_{16}$N$_2$}
\author{Liqin Ke}
\affiliation{Ames Laboratory US DOE, Ames, Iowa 50011}
\author{Kirill D. Belashchenko}
\affiliation{Department of Physics and Astronomy and Nebraska Center for Materials and
Nanoscience, University of Nebraska-Lincoln, Lincoln, Nebraska 68588}
\author{Mark van Schilfgaarde}
\affiliation{Department of Physics, King's College London, Strand, London WC2R 2LS,
United Kingdom}
\author{Takao Kotani}
\affiliation{Tottori University, Tottori, Japan}
\author{Vladimir P Antropov}
\affiliation{Ames Laboratory US DOE, Ames, Iowa 50011}

\begin{abstract}
  The electronic structure and magnetic properties of pure and doped
  {\fen} systems have been studied in the local-density (LDA) and
  quasiparticle self-consistent {\gw} approximations.  The {\gw}
  magnetic moment of pure {\fen} is somewhat larger compared to LDA
  but not anomalously large.  The effects of doping on magnetic moment
  and exchange coupling were analyzed using the coherent potential
  approximation. Our lowest estimate of the Curie temperature in pure
  {\fen} is significantly higher than the measured value, which we
  mainly attribute to the quality of available samples and the
  interpretation of experimental results. We found that different Fe
  sites contribute very differently to the magnetocrystalline
  anisotropy energy (MAE), which offers a way to increase the MAE by small
  site-specific doping of Co or Ti for Fe. The MAE also increases under
  tetragonal strain.
\end{abstract}

\eid{identifier}
\date{\today}
\maketitle



\section{Introduction}


Ordered nitrogen martensite {\afen} was first synthesized in bulk form
by quenching of the cubic nitrogen austenite $\gamma$-FeN with a
subsequent annealing. \cite{jack.prsl1951} Quenching initially
produces disordered $\alpha^{\prime}$-FeN, which then orders during
low-temperature annealing to produce {\afen}.  The latter is a
metastable phase with a distorted body-centered tetragonal structure,
which decomposes into $\alpha$-Fe and Fe$_{4}$N near 500 K.


Interest in {\afen} was revived much later when it was synthesized in
thin film form and a very large value ($\sim$3{\mub}) for the average
Fe magnetic moment was reported.\cite{kim.apl1972} This result was not
independently confirmed until twenty years
later. \cite{sugita.jap1991} Owing to the rapid development of the
magnetic recording technologies, this confirmation inspired numerous
studies of thin-film samples.  However, the existence of the ``giant''
Fe moment remains controversial as many researchers did not reproduce
these findings, while others confirmed
them. \cite{cadogan.ajp1997,coey.jap1994,takahashi.jmmm2000}.  The
lack of consistent and reproducible experimental results may be
attributed to the difficulties associated with the preparation of
single-crystal {\fen} and stabilization of nitrogen, as well as with
the accurate measurement of the magnetization in multi-phase Fe
nitride samples.  This issue has recently attracted additional
interest due to the search for new permanent magnetic materials
without rare-earth elements.\cite{coey.scma2012} A new way to prepare
single-phase {\fen} powder was recently reported, along with evidence
of high maximum energy product ($BH_{\text{max}}$). \cite{ccmcef}


Most theoretical studies of the magnetization of {\afen} were
performed using the local density approximation, generalized gradient
approximation (GGA) or {\ldau}, though recently Sims {\etal}
\cite{sims.prb2012} applied a hybrid functional and the {\gw}
approximation to this material.  In LDA or GGA the magnetic moment of
{\fen} is only slightly enhanced compared to elemental Fe. Lai
{\etal}\cite {lai.jpcm1994} included electronic correlations within
{\ldau} and found an enhanced magnetization
{\mscell}=2.85{\mub/Fe}. Wang
{\etal}\cite{wang.ieee2012,ji.njp2010,ji.apl2011} identified a
localized Fe state coexisting with the itinerant states in X-ray
magnetic circular dichroism (XMCD) measurements. They introduced a
specific charge transfer between different Fe sites and obtained a
large {\mave} in {\ldau}.  However, the choice of the correlated
orbitals and the associated value of the Hubbard $U$ parameter is not
well-defined for metallic systems.  For example, the on-site
interaction parameters obtained by Sims {\etal}\cite{sims.prb2012}
using the constrained random phase approximation (RPA) differ
substantially from those proposed by Wang {\etal}.  The quasiparticle
self-consistent {\gw} approximation ({\qsgw})
\cite{vanschilfgaarde.prl2006,kotani.prb2007} is more reliable and
provides a more satisfactory way to determine the ground state density
and magnetic moment. In the present paper we apply this method to
{\fen}.

Studies of exchange interaction, Curie temperature ({\tc}), and MAE of
{\fen} met with additional difficulties. In particular, measurements
of {\tc} are hampered by the decomposition of the metastable {\fen}
into Fe$_{4}$N and Fe, which was reported to occur above 200{\deg},
\cite{takahashi.jmmm2000} in the $230$-$300${\deg} range,
\cite{kim.apl1972} or at $400${\deg}.\cite{sugita.jap1991} Sugita
     {\etal} extrapolated their data to estimate {\tc} at
     $540${\deg}. \cite{sugita.jap1991} Thermal stability of {\fen}
     was reported to increase with addition of Co and Ti
     \cite{jiang.jap1999,wang.jpd1997} (up to 700{\deg} in the Ti
     case). However, no experimental information is presently
     available about the {\tc} of Co or Ti-doped {\fen}, or of any
     other {\fen} samples stabilized at high temperatures. To the best
     of our knowledge, there have been no theoretical studies of the
     exchange interaction and Curie temperature in {\fen}. Systematic
     studies of the effects of doping on $M$ and {\tc} in {\fen} also
     appear to be lacking.

As for the MAE, only a few experimental values were reported, and they
are varied and inconclusive. For example, Sugita {\etal}
\cite{sugita.jap1991} obtained an in-plane MAE, while
Takahashi\cite{takahashi.ieee1999} found a large uniaxial MAE. The
only available theoretical calculations of MAE used an empirical tight
binding (TB) approximation. \cite{uchida.jmmm2007}


In this paper we study the magnetization, Curie temperature, and
magnetocrystalline anisotropy energy of pure and doped {\fen} using
several well-tested electronic structure techniques and suggest
possible routes for improving its properties for permanent magnet
applications.

\section{Computational methods}


Most LDA, GGA and {\qsgw} calculations were performed using a
full-potential generalization\cite{methfessel.chap2000} of the
standard linear muffin-tin orbital (LMTO) basis
set.\cite{andersen.prb1975} This scheme employs generalized Hankel
functions as the envelope functions.  Calculations of MAE we also
performed using the recently-developed mixed-basis full-potential
method,\cite{kotani.prb2010} which employs a combination of augmented
plane waves and generalized muffin-tin orbitals to represent the wave
functions. The results of a traditional non-self-consistent
application of the {\gw} approximation depend on the non-interacting
Hamiltonian generating the self-energy.  This issue can be
particularly problematic for metals. In contrast, {\qsgw} method does
not suffer from this limitation: it is more reliable than the standard
$GW$. This method gives quasiparticle energies, spin moments,
dielectric functions, and a host of other properties in good agreement
with experiments for a wide range of materials, including correlated
ones such as NiO.  The details of {\qsgw}
implementation\cite{vanschilfgaarde.prl2006,kotani.prb2007} and
applications can be found elsewhere.


The pair exchange parameters were obtained using two linear response approaches:

(1) Static linear-response approach \cite{liechtenstein.jmmm1987}
implemented within the atomic sphere approximation (ASA) to the
Green's function (GF) LMTO method.  \cite{vanschilfgaarde.jap1999} In
addition to making a spherical approximation for the potential, this
method makes the long-wave approximation (LWA), so that the pair exchange
parameter is proportional to the corresponding spin susceptibility
$\chi _{ij}$ \cite{a1}.  The exchange parameters ${A_{ij}}$ obtained
in this method are related to the parameters of the classical
Heisenberg model
\begin{equation}
\label{eq:hsbg}
{H=-\sum_{ij}J_{ij}\mathbf{S}_{i}\cdot \mathbf{S}_{j},}
\end{equation}
by the following renormalization for ferromagnetic(FM) and
antiferromagnetic(AFM) cases:


\begin{align}
  J_{ij} &= A_{ij}/\mathbf{S}_{i}\mathbf{S}_{j} 
   \\
  & =4A_{ij}/\mathbf{m}_{i}\mathbf{m}_{j}= \left\{
\begin{array}{rl}
{4A_{ij}/m_{i}m_{j}} & \text{FM } \\
 {-4A_{ij}/m_{i}m_{j}} & \text{AFM } \\
\end{array} \nonumber \right.
\end{align}
where $\mathbf{m}_{i}$ is the magnetic moment on site {$i$}. With this
renormalization all results obtained for the Heisenberg model
Eq.\eqref{eq:hsbg} can be used directly. Thus, parameters $A_{ij}$
always stabilize (destabilize) the given magnetic configuration and
can be treated as stability parameters.  Curie temperature in the spin
classical mean field approximation (MFA) is simply {\tc}=$2/3\sum_{ij}
A_{ij}$.
 
(2) Dynamical linear response approach with the bare susceptibility
$\chi ( \mathbf{q},\omega )$ calculated in the full product basis set
representation using the LDA or {\qsgw} electronic
structure.\cite{kotani.jpcm2008}.  The results are then projected onto
the functions representing local spin densities on each magnetic site,
which gives a matrix $\chi _{ij}( \mathbf{q},\omega )$ in basis site
indices.\cite{kotani.jpcm2008} This projection corresponds to the
rigid spin approximation. The inversion of this matrix with a
subsequent Fourier transform provides the real-space representation of
the inverse susceptibility representing the effective pair exchange
parameters:
\begin{equation}
J_{ij}=\underset{\omega \longrightarrow 0}{\lim }\frac{1}{\Omega _{BZ}}\int d%
\mathbf{q}\left[ \chi (\mathbf{q},\omega )\right] ^{-1}e^{i\mathbf{qR}_{ij}}.
\end{equation}


{\tc} is calculated both in the MFA\cite{anderson.ssp1963} and the
RPA-Tiablikov\cite{rusz.prb2005} approximations.
The actual {\tc}  may usually be expected to lie between the results of these two approximations.


To address the effects of doping, we used our implementation of the
coherent potential approximation (CPA) within the TB-LMTO code, which
follows the formulation of Turek {\etal}\cite{turek97} and
Kudrnovsk\'y {\etal}.\cite{kudrnovsky.prb1990} A coherent interactor
matrix $\Omega _{i}$ is introduced for each basis site $i$ treated
within CPA. At self-consistency $g_{ii}=(\EuScript{P}_{i}-\Omega
_{i})^{-1}$, where $\EuScript{P}_{i}$ is the coherent potential matrix
for site $i$, and $g_{ii}$ is the on-site block of the average
auxiliary LMTO GF matrix $g=(\EuScript{P}-S)^{-1}$ . This on-site
block is extracted from the Brillouin zone integral of
$g(\mathbf{k})$. The conditionally averaged GF at site $i$ occupied by
component $a$ is $g_{ii}^{a}=(P_{a}-\Omega )^{-1}$, and the CPA
self-consistency condition can be written as
$g_{ii}=\sum_{a}c_{i}^{a}g_{ii}^{a}$; here $c_{i}^{a}$ is the
concentration of component $a$ at site $i$. Using this equation, at
the beginning of each iteration the stored matrices $\Omega _{i}$ are
used to obtain an initial approximation to $\EuScript{P}_{i}$. In
turn, $\EuScript{P}_{i}$ is used in the calculation of $g_{ii}$ by a
Brillouin zone integral. The next approximation for $\Omega _{i}$ is
obtained from $\Omega _{i}=\EuScript{P}_{i}-g_{ii}^{-1}$. These output
matrices are then linearly mixed with the input $\Omega_{i}$ matrices
at the end of the iteration.

We found that the mixing coefficient of 0.4 for $\Omega_i$ works well
in most cases.  For fastest overall convergence, we found that it is
usually desirable to iterate CPA iterations until the $\Omega_{i}$
matrices are converged to a small tolerance, and only then perform the
charge iteration. The convergence of $\Omega$ is done separately for
each point on the complex contour to the same tolerance. With this
procedure, fairly aggressive Broyden mixing can be used for LMTO
charge moments. CPA convergence at each charge iteration usually takes
10-50 iterations depending on the imaginary part of energy and the
selected tolerance. At the beginning of the calculation, the
$\Omega_{i}$ matrices are set to zero; afterwards they are stored and
reused for subsequent iterations. In order to avoid unphysical
symmetry-breaking CPA solutions (which otherwise often appear), the
coherent potentials and the $\mathbf{k}$-integrated average auxiliary
GF are explicitly symmetrized using the full space group
of the crystal. As a result, the use of CPA does not impose any
restrictions on the symmetry of the crystal. Calculations reported
here were performed without using charge screening corrections for the
Madelung potentials and total energy.

The effective exchange coupling in CPA is calculated as
\begin{equation}
A_{0}\left( c\right) =cA_{X}\left( c\right) +(1-c)A_{Y}\left( c\right)
\end{equation}
where the component-specific $A_{i}\left( c\right) $ are obtained
using the conditionally averaged GF and the formalism of
Ref.\ \onlinecite{a3}.


For MAE calculations the self-consistent solutions are found including
spin-orbit coupling (SOC) terms of order $1/c^{2}$.  The MAE is
defined below as $K=E_{100}{-}E_{001}$, where $E_{001}$ and $E_{100}$
are the total energies for the magnetization oriented along the
$[001]$ and $[100]$ directions, respectively. Positive (negative) $K$
corresponds to uniaxial (planar) anisotropy.  We used a $24\times
24\times 24$ $k$-point mesh for MAE calculations to ensure sufficient
convergence; MAE changed by less than 2\% when a denser $32\times
32\times 32$ mesh was employed. All calculations except {\qsgw} were
performed with both LDA \cite{vonbarth.jpcssp1972} and GGA
\cite{perdew.prl1996} exchange-correlation potentials for comparison.


\section{Results and discussion}

\begin{figure}[tbp]
\begin{tabular}{c}
\includegraphics[width=.50\textwidth,clip,angle=0]{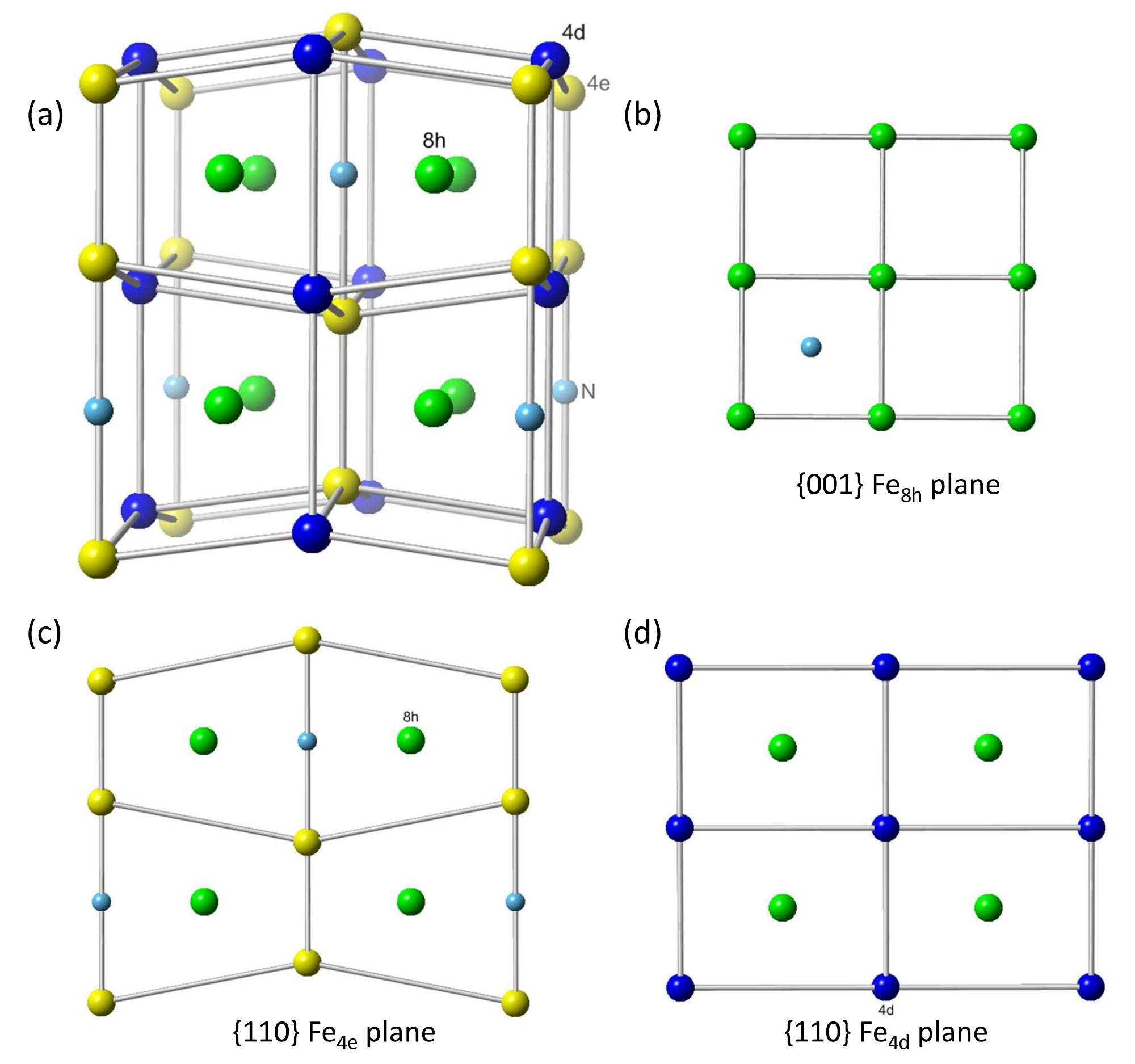}
\end{tabular}
\caption{ (Color online) ({a}) Crystal structure of {\fen}. The
  experimental atomic positions are shown. Relaxed structure have
  slightly different $z_{4e}$ and $x_{8h}$. ({b}) $\{001\}$~plane with
  {${8h}$} and N atoms. ({c}) $\{110\}$~plane with ${4e}$, {${8h}$}
  and N atoms. ({d}) $\{110\}$~plane with ${4d}$ and {${8h}$} atoms. }
\label{fig:xtal}
\end{figure}

The crystal structure of {\fen} is body-center-tetragonal (bct) with
space group $I_{4/mmm}$($\#139$).  It may be viewed as a distorted
2$\times $2$\times $2 bct-Fe superlattice with $c/a$=1.1.  Crystal
structure of {\fen} was first identified by Jack.\cite{jack.prsl1951}
Here we use lattice constants $a$=$b$=$5.72$\AA\, $c$=$6.29${\AA } and
atomic position parameters $z_{4e}$=$0.3125$ and $x_{8h}$=$0.25$ from
Jack's work as the experimental structure.  (see Fig.~\ref{fig:xtal}).

We also relaxed the structure by minimizing the total energy in LDA
and obtained $z_{4e}$=$0.293$ and $x_{8h}$=$0.242$, nearly identical to
that obtained by Sawada {\etal}\cite{sawada.prb1994}. The primitive
cell contains one N and eight Fe atoms divided into three groups
indicated by Wyckoff sites: two {${4e}$}, four {${8h}$} and two
{${4d}$} sites (correspondingly first, second and third neighbors to
N).

\subsection{Magnetic moments and electronic structure}

Table~\ref{tbl:qm} shows the atomic spin moment {\mspin} at the three
Fe sites and magnetization {\mscell}(orbital magnetic moment is small,
hereafter we only include spin magnetization in {\mscell}). Within the
LDA, {\mscell}=$2.38${\mub/Fe} was obtained, in good agreement with
previously reported calculations
\cite{cadogan.ajp1997,coey.jap1994,takahashi.jmmm2000}. The
enhancement relative to elemental bcc-Fe has been attributed to the
size effect\cite{martar.zpb1992}. {\qsgw} gives
{\mscell}=$2.59${\mub/Fe}, about $9\%$ larger than LDA.  While it is
known that {\gw} enhances spin moments relative to LDA, to ensure the
genuineness of this enhancement of moment, we also carried out the
{\qsgw}\ calculation of bcc-Fe and found that {\qsgw} enhance the LDA
magnetic moment in elemental Fe by only ${\sim }2$\%
(2.20$\rightarrow$2.24{\mub}). Sims {\etal}\cite{sims.prb2012} found
a similar {\mscell} in their {\gw} calculation while they also
obtained a larger magnetization enhancement in bcc-Fe with
{\mscell}=2.65{\mub/Fe}. Considering {\fen} consists of about $87\%$
Fe, the $9\%$ enhancement we find non-trivial.  However, it is still
well below {\mscell$=2.85${\mub/Fe}}, obtained in {\ldau} by Lai
{\etal}\cite{lai.jpcm1994}. The spin moment on the {$4d$} site reaches
{\mspin}$=3.11${\mub} in {\qsgw}, though we do not observe any obvious
charge transfer from {${4d}$} to {${4e }$} and {${8h}$} sites in
{\qsgw}, relative to the LDA.  Hence, we can not attribute the
enhancement of {\mscell} to the charge transfer as suggested by others
\cite{wang.ieee2012,ji.njp2010,ji.apl2011}.

\begin{table}[tbp]
\caption{Atomic spin magnetic moment {\mspin} and spin magnetization
  {\mscell} in {\fen} in different methods. Calculations are in the
  LDA unless GGA or {\qsgw} is specified. }
\label{tbl:qm}
\begin{tabular}{l|c|c|c|c|c|c|c}
\hline
\multirow{2}{*}{Method} & \multicolumn{4}{c|}{{\mspin}(\mub)\footnotemark[1]} &{\msave}\footnotemark[2] & \multicolumn{2}{c}{{\mscell}\footnotemark[3]}   \\ \cline{2-5} \cline{7-8}
              & $4e$  & $8h$  & $4d$  &  N    & {(\mub)} &  {(\mub/Fe)}   & (emu/g) \\  \hline
ASA           &  2.07 &  2.40 &  3.03 & -0.06 &  2.48 &  2.47  &   239    \\
ASA-GF        &  2.10 &  2.41 &  2.99 & -0.10 &  2.48 &  2.47  &   239    \\
FP            &  2.08 &  2.32 &  2.84 & -0.05 &  2.39 &  2.38  &   231    \\
FP(GGA)       &  2.21 &  2.40 &  2.86 & -0.04 &  2.47 &  2.43  &   236    \\
{\qsgw}       &  2.24 &  2.55 &  3.12 & -0.01 &  2.62 &  2.59  &   251    \\ \hline
\end{tabular}\\ 
\footnotetext[1]{Spin moment inside atomic or muffin-tin sphere.}
\footnotetext[2]{Average of the atomic spin moments of all Fe sites
  without taking account of interstitial and N sites.}
\footnotetext[3]{Average spin moments within the cell (with taking
  account of interstitial and N sites).}
\end{table}

\begin{figure}[ht]
\centering
\par
\begin{tabular}{cc}
\includegraphics[width=.46\textwidth,clip]{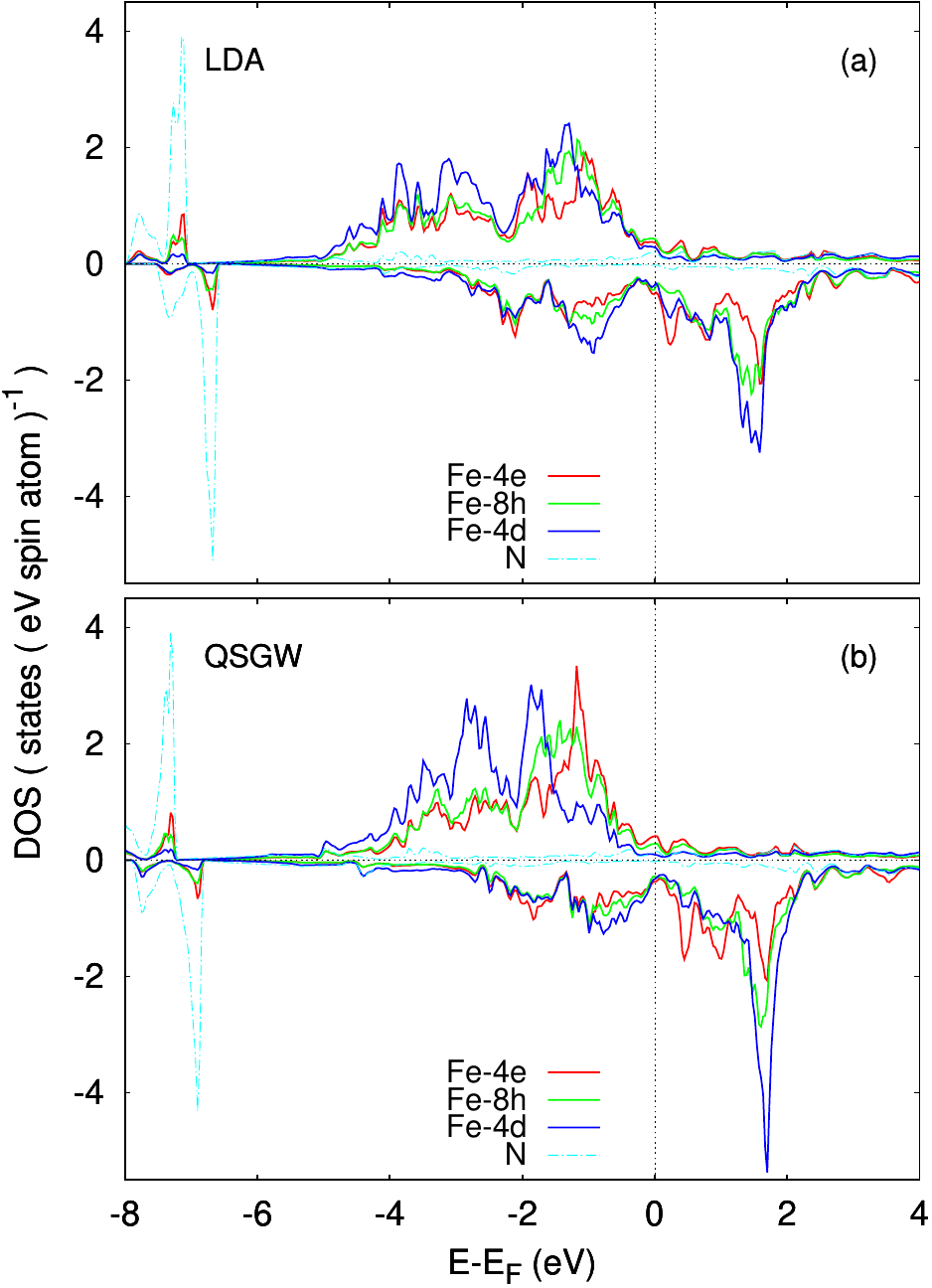} & 
\end{tabular}%
\caption{(Color online) Site and spin-projected densities of states
  within LDA (a) and {\qsgw} (b). }
\label{fig:dos}
\end{figure}


Density of states (DOS) calculated within LDA and {\qsgw} are shown in
Fig.~\ref{fig:dos}.  The LDA result is similar to previously
reported results.  A careful examination of the band structure reveals that
{\qsgw} significantly modifies the energy bands near
{$E_{\mathrm{F}}$}, relative to LDA.  It has a slightly larger on-site
exchange, widening the split between the majority and minority DOS and
increasing {\mave} by about $9\%$. Both DOS figures show hybridization
between N-$2p$ and Fe-$3d$ states at around {$-7$}~eV, indicating that
{\qsgw} does not strongly modify the relative alignment of N-$2p$ and
Fe-$3d$ levels. Comparing the partial DOS reveals that bands are
slightly wider and hybridization is overestimated in LDA, as is
typical since LDA tends to overestimate $3d$ bandwidths slightly. The
DOS also show hybridization is stronger in the {${4e}$} and {${8h}$}
channels while weaker in the {${4d}$} channels, which are the furthest
removed from N. Also, as typical with second row elements, {\qsgw}
pushes the N-$2s$ bands down relative to LDA, from -16.2~eV to -18~eV.
The N-$2s$ also hybridizes with {Fe-${4e}$} and {Fe-${8h}$}. However
there is almost no hybridization with the furthest {Fe-${4d}$} at all
because the N-$2s$ orbitals is very localized.

\subsection{Exchange coupling and Curie temperature}

\begin{figure}[tbp]
\begin{tabular}{c}
\includegraphics[width=.49\textwidth,clip,angle=0]{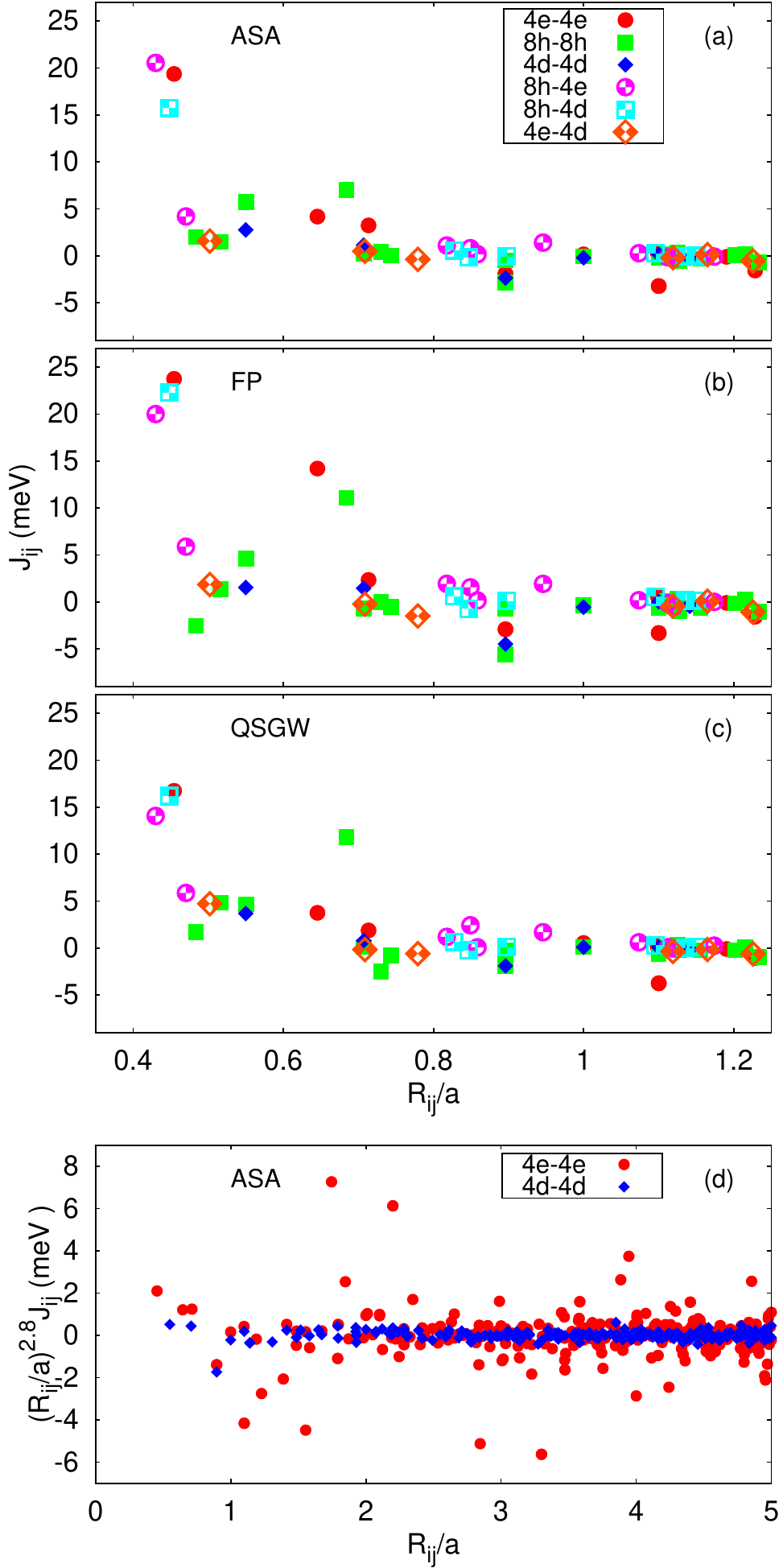} 
\end{tabular}
\caption{ (Color online) Real-space magnetic exchange parameters
  {\jij} in {\fen} within ASA-GF (a), FP (b), and {\qsgw} (c) as
  functions of distance $R_{ij}/a$. (d) $(R_{ij}/a)^{2.8}J_{ij}$ in
  ASA-GF as a functions of distance $R_{ij}/a$.  The in-plane lattice
  constant $a$ in {\fen} is as twice large as in bcc-Fe.}
\label{fig:jr}
\end{figure}

\begin{table}[tbp]
\caption{Pairwise exchange parameters of the Heisenberg model {\jij}(meV) and {\tc}(K)
  calculated with different methods.}
\label{tbl:exchange}%
\begin{tabular}{c|c|ccc|c|c|c}
\hline
& $|R_{ij}|/a$ &  & direction &  & ASA & FP & {\gw} \\ 
\hline
\fee-\fee    
    & 0.454 &       0 &       0 &  -0.455  &  19.37  &   23.75  &  16.75  \\
    & 0.645 &       0 &       0 &   0.645  &   4.18  &   14.20  &   3.76  \\
    & 0.713 &  -0.5   &  -0.5   &   0.095  &   3.24  &    2.33  &   1.88  \\
    & 0.895 &  -0.5   &  -0.5   &  -0.550  &  -1.92  &   -2.91  &  -1.44  \\
\hline                                                                    
\feh-\feh   
    & 0.483 &       0 &  -0.484 &       0  &   2.00  &   -2.55  &   1.69  \\ 
    & 0.516 &       0 &   0.516 &       0  &   1.49  &    1.37  &   4.84  \\
    & 0.550 &   0.016 &   0.016 &   0.550  &   5.74  &    4.60  &   4.67  \\
    & 0.684 &  -0.484 &  -0.484 &       0  &   7.00  &   11.07  &  11.80  \\
    & 0.707 &   0.516 &  -0.484 &       0  &   0.20  &   -0.68  &   0.13  \\
    & 0.730 &   0.516 &   0.516 &       0  &   0.39  &   -0.01  &  -2.50  \\
\hline                                                                    
\fed-\fed   
    & 0.550 &       0 &       0 &   0.550  &   2.76  &    1.54  &   3.67  \\
    & 0.707 &  -0.5   &   0.5   &       0  &   1.15  &    1.44  &   0.78  \\
    & 0.895 &  -0.5   &  -0.5   &   0.550  &  -2.37  &   -4.49  &  -1.90  \\
\hline                                                                   
\fee-\feh   
    & 0.430 &  -0.258 &  -0.258 &  -0.227  &  20.55  &   20.00  &  14.08 \\
    & 0.470 &   0.242 &   0.242 &   0.323  &   4.20  &    5.88  &   5.88 \\
\hline                                                                    
\feh-\fed   
    & 0.448 &   0.258 &  -0.242 &   0.275  &  15.73  &   22.30  &  16.18  \\
    & 0.827 &  -0.242 &  -0.742 &  -0.275  &   0.58  &    0.60  &   0.64  \\
\hline                                                                    
\fed-\fee   
    & 0.502 &  -0.5   &       0 &   0.048  &   1.56  &    1.85  &   4.72  \\
    & 0.708 &       0 &  -0.5   &  -0.502  &   0.49  &   -0.23  &  -0.16  \\
\hline
\multicolumn{3}{c}{{\tc}(MFA)}   &         &          &   1552  & 1621     &    1840  \\
\multicolumn{3}{c}{{\tc}(RPA)}   &         &          &   1118  & 1065     &    1374  \\
\hline
\end{tabular}%
\end{table}

The Heisenberg model parameters {\jij} using LDA-ASA in the LWA,
FP-LDA and {FP-\qsgw} are plotted in Fig.~\ref {fig:jr} and tabulated
in Table~\ref{tbl:exchange}.  The two LDA results are quite similar,
confirming that the ASA and the LWA form a reasonable approximation.
{\qsgw} shows some differences, particularly reducing those
interactions which are AFM.

The structure of {\jij} is much more complicated in {\fen} than in
elemental bcc-Fe. The vectors connected the nearest {${8h}$}-{${4e}$}
or {${8h}$}-{${4d}$} sites are nearly along $[111]$ direction, the
magnetic interactions between them are generally large. In comparison,
the largest interaction is also between the nearest sites connected by
vectors along the $[111]$ direction in bcc-Fe.
{\jij} between {${8h}$}-{${4e}$} sites is very anisotropic due to the
distortion of lattice around N atom. Similar anisotropy was also found
for the in-plane couplings on the {${8h}$} lattice.  Interestingly, a
large coupling, ({\jij}=23.75~meV in the FP-LDA calculation), occurs
between two {${4e}$} atoms along $[001]$. This pair has been squeezed
together by neighboring N atoms. Since exchange coupling is sensitive
to the distance between those two sites, we also examined this
exchange parameter for the experimental atom coordinates, for which
the bond length of the {${4e }$}-{${4e}$} pair shrinks from {2.60} to
{2.36} \AA , and found that this {\jij} increases from {23.75} to
{36.55} {meV}, indicating significant exchange-striction effect. The
second nearest {${4e}$}-{${4e}$} coupling(two Fe atoms with a N atom
between them along $\langle 110\rangle $ direction) is {14.2~meV} in
FP and 4.18 {meV} in ASA. The relatively large disagreement may be a
consequence of the shape approximation used in ASA, considering the
presence of N atom and strong lattice distortion around this pair of
atoms.

The calculated magnetic interactions between different types of atoms
have very different spatial dependence and correspondent asymptotic
behavior. To demonstrate it explicitly on Fig.~\ref{fig:jr}(d) we show
{\jij} scaled with $(R_{ij}/a)^{2.8}$. With this renormalization
{\jij} between atoms on 4e positions (smaller moments) are
approximately constant in this range of distances(long-ranged
interaction), while {\jij} between Fe atoms on 4d sites (with the
largest moments) decay much faster( short-ranged interaction),
corresponding to more localized moment behavior. Such very different
asymptotic behaviour suggests that these localized and delocalized
interactions correspond to Fermi surface shapes with different
dimensionalities.

{\tc} is calculated from the exchange parameters and tabulated in
Table~\ref{tbl:exchange}. RPA values are about {$30\%$} smaller than
the MFA ones. Typically experimental values fall between the MFA and
RPA results, with the RPA being closer to the experiment in normal
three dimensional systems. In the present case, however, the reported
extrapolated experimental estimate {{\tc}=810}K \cite {sugita.jap1991}
is smaller than both of MFA and RPA values, and smaller than the one
in bcc-Fe\cite{vanschilfgaarde.jap1999}.  This is rather unusual.  In
bcc-Fe, we estimate {\tc} to be $\sim$1300 K and $\sim$900 K in the
MFA and RPA respectively, which bracket the experimental value of 1023
K (as is typical).  Contrary to experiment, our calculated {\tc} for
{\fen} is \emph{higher} than for bcc Fe in all our estimations.  Such
disagreement between theory and experiment is much larger in {\fen}
than other Fe-rich phases.  The disagreement may originate from
approximations to the theory (absence of spin quantum effects,
temperature-dependence of exchange, among others) that uniquely affect
{\fen}; or alternatively from the experimental interpretation of the
measured {\tc}.  We cannot completely discount the former possibility,
but for this local-moment system, it is unlikely that the most serious
errors originate in density functional theory that generate exchange
parameters.  For instance, parameters generated from QS\emph{GW} also
do not improve agreement with the experiment; indeed this increase the
discrepancy.  On the other hand, as noted in the introduction {\fen}
decomposes with increasing $T$; moreover, there is a transition to the
(nonmagnetic) {$\gamma$} phase at 1185 K. Since the measured $M(T)$ is
not a measurement of the single-phase material, it is still unknown
what is the experimental value for the {\tc} in a single-phase {\fen}.
Unfortunately, disentangling the structural and magnetic degrees of
freedom is very difficult, experimentally. Finally, if {\mave}
increases in {\fen}, as is observed and predicted, {\tc} should
increase. Thus we conclude that {\tc} in high quality samples of pure
{\fen} is probably larger than what has been reported so far, and
larger than in pure bcc-Fe.


Let us now discuss the influence on {\tc} when other atoms 
substitute for Fe or N.

The CPA is an elegant, single-site \emph{ab initio} approach to study
substitutional alloys.  We have implemented the CPA, including a MFA
estimate for {\tc}, within the ASA.  As we have seen by comparing
exchange interactions in the ASA to those without this approximation,
the ASA does not seem to be a serious approximation to the LDA in this
material.

\begin{figure}[tbp]
\begin{tabular}{c}
\includegraphics[width=.49\textwidth,clip,angle=0]{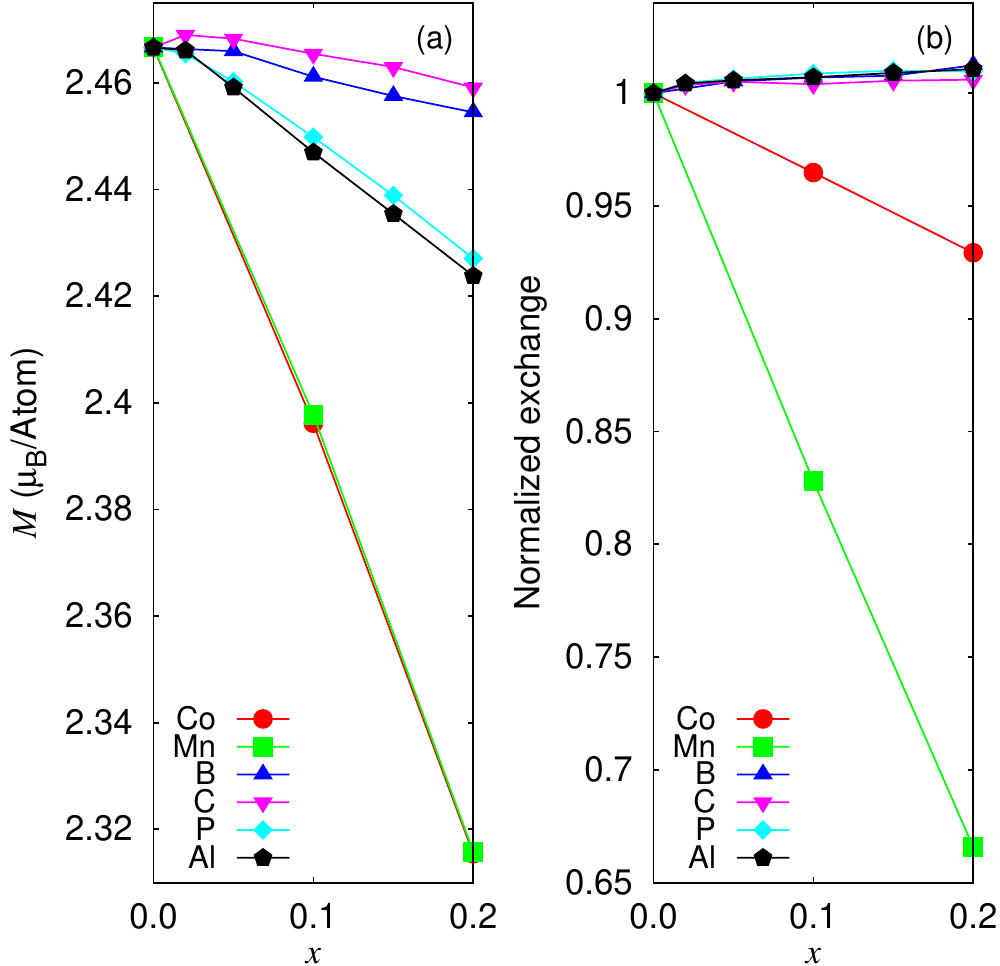} \\ 
\end{tabular}
\caption{ (Color online) Spin magnetization {\mscell} (a) and
  normalized exchange {$J_{0}/J_{0}$}({\fen}) (with respect to 
  pure {\fen}) (b) as functions of doping concentration in {\fen}. The
  concentration {$x$} of doping element {T} is defined as
  (Fe$_{1-x}$T$_{x}$)$_{16}$N$_2$ with Fe site doping (T=Co,Mn); and
  Fe$_{16}$(N$_{1-x}$T$_{x}$)$_2 $ with N site doping(T=B,C,P and Al)
  . }
\label{fig:jrcpa}
\end{figure}

Fig.~\ref{fig:jrcpa} shows the {\mave} and and the normalized
effective exchange (or MFA estimation of {\tc} in units of pure
{\fen}) with doping by different elements. Both Co and Mn doping cause
{\tc} to decrease. On the other hand, with N site being doped with
B,C,P or Al elements, {\tc} and {\mave} change slightly and the Fermi
surface character is barely affected.

\begin{figure}[tbp]
\begin{tabular}{c}
\includegraphics[width=.48\textwidth,clip,angle=0]{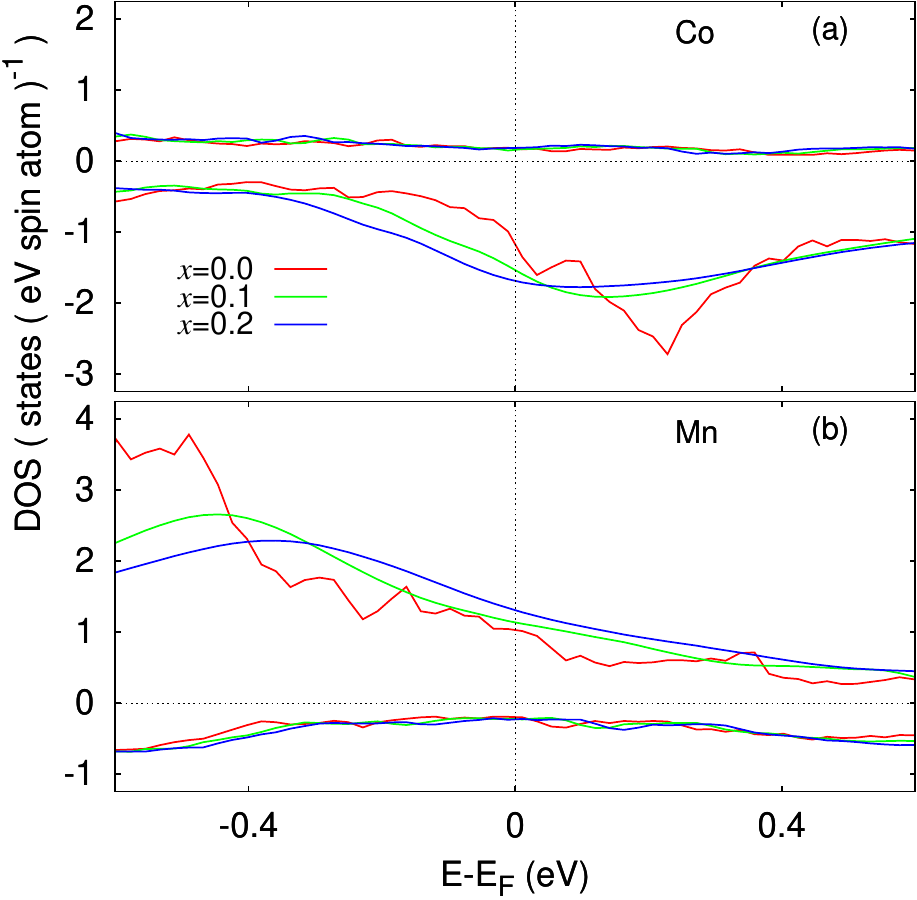}
\\
\end{tabular}%
\caption{ (Color online) Densities of state of substitutional
  component in random alloy (Fe$_{1-x}$Co$_{x}$)$_{16}$N$_2$ (a) and
  (Fe$_{1-x}$Mn$_{x}$)$_{16}$N$_2$ (b).}
\label{fig:pdoscpa}
\end{figure}

\begin{table*}[ht]
\caption{Component-resolved atomic spin moments {\mspin} ( the
  atomic spin moment of substitutional component are given in parentheses),
  magnetization {\mscell} and exchanges $J_0$ in Co and Mn-doped
  {\fen} calculated within ASA-GF. }
\label{tbl:cpamj0}%
\begin{tabular}{|c|c|cc|cc|cc|c|cc|cc|cc|}
\hline
\multirow{2}{*}{Substituent} & \multirow{2}{*}{$x$}  &   \multicolumn{6}{c|}{{\mspin}(\mub)} & \multirow{2}{*}{\mscell(\mub/atom)} & \multicolumn{6}{c|}{$J_{0}(meV)$} \\  \cline{3-8}  \cline{10-15}
                  &     & \multicolumn{2}{c|}{${4e}$}  & \multicolumn{2}{c|}{${8h}$} & \multicolumn{2}{c|}{${4d}$} &  &  \multicolumn{2}{c|}{${4e}$}    & \multicolumn{2}{c|}{${8h}$}    &  \multicolumn{2}{c|}{${4d}$}   \\ \hline
\multirow{3}{*}{Co}  & 0.00 &  2.10 & ( 1.44) &  2.41 & ( 1.68) &  2.99 & ( 2.11) &  2.47 & 12.95 & (11.37) & 15.70 & (14.62) & 16.96 & (19.42) \\
                     & 0.10 &  2.09 & ( 1.34) &  2.43 & ( 1.64) &  3.01 & ( 2.09) &  2.40 & 11.97 & ( 9.62) & 15.45 & (13.56) & 16.70 & (18.68) \\
                     & 0.20 &  2.08 & ( 1.27) &  2.45 & ( 1.62) &  3.02 & ( 2.08) &  2.32 & 11.24 & ( 8.47) & 15.19 & (12.91) & 16.50 & (18.07) \\  \hline
\multirow{3}{*}{Mn}  & 0.00 &  2.10 & ( 1.88) &  2.41 & ( 2.25) &  2.99 & ( 3.23) &  2.47 & 12.96 & ( 7.05) & 15.70 & ( 6.57) & 16.97 & ( 1.07) \\
                     & 0.10 &  2.07 & ( 1.75) &  2.35 & ( 2.04) &  2.96 & ( 3.03) &  2.40 & 12.03 & ( 5.64) & 14.01 & ( 4.03) & 14.97 & (-1.10) \\
                     & 0.20 &  2.06 & ( 1.64) &  2.30 & ( 1.85) &  2.93 & ( 2.90) &  2.32 & 11.15 & ( 4.41) & 12.59 & ( 2.09) & 13.22 & (-2.69) \\
\hline
\end{tabular}%
\end{table*}

As shown in Fig.~\ref{fig:jrcpa}, Co or Mn-doped {\fen} decrease
moment and exchange coupling.  We neglected the possible site
preference effect in this calculation, and doped all three Fe sites
with equal probability.  Table~\ref{tbl:cpamj0} shows the magnetic
moment and $J_{0}$ parameters of Fe and substitutional components on
all three different Fe sites. It indicates an opportunity to increase
{\tc} by using a separate Co-doping on Fe-$4d$ sites.

Magnetic moments of the Fe component decrease with Mn doping and
slightly increase with Co doping.  With Co doping, the Fe moments on
{$8h$} and {$4d$} sites do slightly increase, however this increment
is not big enough to overcome the decrease resulting from Co
substituting for Fe - the system behaves more like localized moments
system. We also carried out the FP calculation of Fe$_7$CoN, with one
out of eight Fe atoms being replaced by Co atom and confirmed that the
magnetization decrease, especially when Co replace the Fe on {\fee}
site. This is consistent with the CPA results. Mn substituent have
larger magnetic moments than Co substituent. However, Mn doping
decreases moments on Fe sites.  Overall, the dependence of the total
magnetic moments on substituent concentration are almost exactly same
with Co and Mn doping. Another interesting observation is that
magnetic moments of both substituents decrease with increasing of
doping concentration.  This can be explained by the partial density of
states as shown in \rfig{fig:pdoscpa}. The magnetic moment of Co
slightly decreases with increasing of doping concentration.  As shown
in \rfig{fig:pdoscpa}, the unoccupied DOS peak right above the Fermi
energy ({\efermi}) in the minority spin channel moves toward it.  More
electrons fill in the minority channel and decrease the magnetic
moment as doping increases.  With Mn doping on the other hand, the
peak in the majority channel right below {\efermi} becomes less
pronounced.  It shifts toward {\efermi} and decreases the magnetic
moment of Mn component. For the Fe component DOS, there is no peak
structure near the {{\efermi}, and magnetic moment is much less
  sensitive to the substitutional concentration.

\subsection{Magnetic anisotropy}

\begin{table}[tbp]
\caption{Previous works on magnetic anisotropy in {\fen}.}
\label{tbl:anisotropy}%
\begin{tabular}{c|c|c|c|c}
\hline
Method                &                     & $K$($\frac{10^5\text{erg}}{\text{cm}^3}$) & Easy axis & Ref. \\ \hline
Exp.                  & Sugita    {\etal}   & 4.8                                       & [100]     & \cite{sugita.jap1991} \\ 
                      & Takahashi {\etal}   & $200$\footnotemark[1]                     & [001]     & \cite{takahashi.ieee1999}\\ 
                      & Takahashi {\etal}   & 97                                        & [001] & \cite{uchida.jmmm2007} \\ 
                      & Kita      {\etal}   & $44$                                      & [001] & \cite{kita.jmmm2007} \\ 
                      & Ji        {\etal}   & $100$\footnotemark[2]                     & [001] & \cite{ji.prb2011} \\ \hline
TB\footnotemark[3]   & Uchida {\etal}      & 140                                       & [001] & \cite{uchida.jmmm2007} \\ 
\hline
\end{tabular}
\footnotetext[1]{Value of ($K_1$+$K_2$).} 
\footnotetext[2]{Measured in partial-ordering {\fen}, author claimed MAE should be much higher for the single-phase sample.} 
\footnotetext[3]{Tight binding approximation}
\end{table}

\begin{table}[tbp]
\caption{The MAE {$K$}, on-site orbital magnetic moment {\morb} and
  the AMAE {\deso} with different spin quantization axis direction in
  pure, Co-doped and Ti-doped {\fen}. Spin quantization axis direction
  {$\mathbf{e}$} are along $[001]$,$[100]$ and $[110]$ directions
  respectively. With the spin along $[100]$ and $[110]$, {\deso} and
  {$K$} values(with respect to $[001]$ direction) directions are
  given.  To estimate {\deso}, $\xi_{i}$=50,70 $m$eV had been used for
  Fe and Co atoms respectively. }
\label{tbl:orbmom}
\begin{tabular}{c|c|c|c|c|c|c|c}
\hline
                      & \multirow{2}{*}{$\mathbf{e}$}& \multicolumn{2}{c|}{$K$}       & \multicolumn{3}{c|}{{\morb}($10^{-3}\mu_B$)}      & {\deso}\\ \cline{3-7}
                      &                           & $\frac{\mu\text{eV}}{\text{Fe}}$  & $\frac{10^5\text{erg}}{\text{cm}^3}$ & $4e$  & $8h$  & $4d$  & $\frac{\mu\text{eV}}{\text{Fe}}$\\ \hline
Exp.\footnotemark[1]  & $001$                   &             &                  & 54                     & 45                           & 71                      &     \\ 
                      & $100$                   & 116         & 144              & 35                     & 49                           & 64                      &  \\
                      & $110$                   & 116         & 144              & 36                     & 58 39                        & 64                      &  \\ \hline
Exp.\footnotemark[1]  & $001$                   &             &                  & 52                     & 44                           & 68                      &  \\
GGA                   & $100$                   & 105         & 131              & 36                     & 48                           & 61                      & 110 \\
                      & $110$                   & 105         & 131              & 36                     & 57 39                        & 61                      & 110 \\ \hline
Theo.\footnotemark[2] & $001$                   &             &                  & 62                     & 46                           & 67                      &     \\
                      & $100$                   & 84          & 103              & 39                     & 50                           & 63                      & 137 \\
                      & $110$                   & 84          & 103              & 39                     & 58 41                        & 63                      & 137 \\ \hline
Theo.\footnotemark[2] & $001$                   &             &                  & 56                     & 43                           & 62                      &  \\
GGA                   & $100$                   & 52          & 65               & 38                     & 47                           & 58                      &  \\
                      & $110$                   & 52          & 65               & 38                     & 55 40                        & 58                      &  \\ \hline \hline

Fe$_7$CoN             & $001$                   &             &                  & 91\footnotemark[4] 72  & 47                           & 76                      &  \\ 
$(4e)$\footnotemark[3]& $100$                   & 165         & 206              & 49\footnotemark[4] 30  & 49                           & 61 64                   & 337 \\ 
                      & $110$                   & 165         & 206              & 49\footnotemark[4] 30  & 61 36                        & 63                      & 336 \\ \hline  
                      & $001$                   &             &                  & 63                     & 69\footnotemark[4] 41 41 44  & 70                      &  \\ 
$(8h)$                & $100$                   & 42          & 52               & 36                     & 77\footnotemark[4] 46 50 47  & 67                      & 123  \\        
                      & $110$                   & 16          & 20               & 40                     & 90\footnotemark[4] 38 38 56  & 67                      &  81  \\ \hline 
                      & $001$                   &             &                  & 63                     & 51                           & 120\footnotemark[4] 81  &  \\  
$(4d)$                & $100$                   & 138         & 171              & 33   38                & 52                           & 106\footnotemark[4] 70  & 271 \\                
                      & $110$                   & 138         & 171              & 36                     & 62 42                        & 106\footnotemark[4] 70  & 271 \\ \hline \hline  

Fe$_7$TiN             & $001$                   &             &                  & 11\footnotemark[4] 63  & 40                           & 69                      &   \\ 
$(4e)$                & $100$                   & 127         & 158              & 10\footnotemark[4] 27  & 46                           & 65 65                   & 62 \\         
                      & $110$                   & 127         & 158              & 10\footnotemark[4] 27  & 55 38                        & 65                      & 62 \\ \hline  
                      & $001$                   &             &                  & 55                     & 14\footnotemark[4] 48 48 41  & 69                      & \\ 
$(8h)$                & $100$                   & 57          & 71               & 38                     & 13\footnotemark[4] 48 50 50  & 65                      & 122\\         
                      & $110$                   & 43          & 53               & 35                     & 13\footnotemark[4] 41 41 62  & 68                      &  83\\ \hline  
                      & $001$                   &             &                  & 60                     & 40                           & 14\footnotemark[4]  69  &   \\ 
$(4d)$                & $100$                   & 102         & 127              & 38   38                & 45                           & 13\footnotemark[4]  67  &  83\\         
                      & $110$                   & 103         & 128              & 38                     & 52 38                        & 13\footnotemark[4]  67  &  83\\ \hline  
\end{tabular}
\footnotetext[1]{Exp. the experimental crystal structure was used.} 
\footnotetext[2]{Theo. the theoretically optimized crystal structure was used. } 
\footnotetext[3]{Doping site of the substitutional atom. }
\footnotetext[4]{Orbital magnetic moments of the substitutional atoms.}
\end{table}

\begin{figure}[tbp]
\begin{tabular}{c}
\includegraphics[width=.48\textwidth,clip,angle=0]{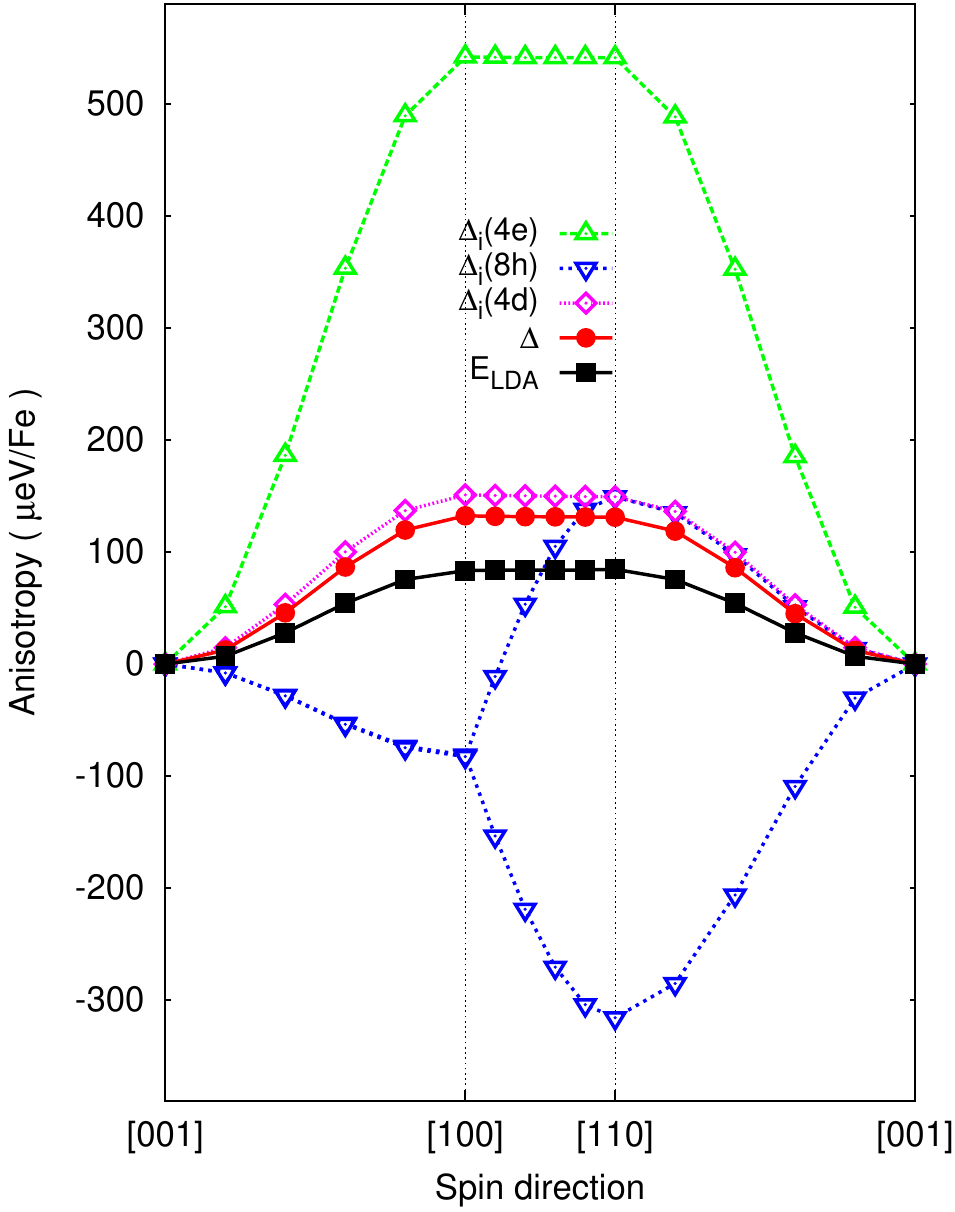}
\end{tabular}
\caption{ (Color online) AMAE on $4e$, $8h$, $4d$ Fe sites ({\desoi})
  and their average value ({\deso}) and the LDA total energy relative
  to the ground state ($E_\text{LDA}$) as functions of spin
  quantization axis rotation. }
\label{fig:kmorb}
\end{figure}

Values of MAE from previous work are summarized in Tables~\ref
{tbl:anisotropy}. Results of present work are shown in
Table~\ref{tbl:orbmom}. All calculations are carried out within LDA
unless GGA is specified. For the pure {\fen}, both experimental and
optimized structure are investigated. Note that doping and SOC lower
the symmetry, and the degeneracy of {\morb} varied.  Within LDA, a
uniaxial magnetic anisotropy $K$=144$\times $ $10^{5}$ erg/cm$^{3}$
was obtained with experimental atomic coordinates. Structural
optimization gives a smaller MAE with $K$= 103${\times
}10^{5}$erg/cm$^{3}$. GGA gives smaller MAE than LDA. It is usually
non-trivial to analyze the origin or site dependence of magnetic
anisotropy. 
Below we define the atomic magnetic anisotropy energy (AMAE)
{$\Delta_{i}$} as half of the difference of the SOC
energies along different magnetic field directions, that is in turn
defined by the corresponding anisotropy of orbital magnetic moments:%
\begin{equation}
\sum \Delta_{i}(\theta=90^{\circ}) =\sum \xi _{i}m_{i}(l_{i}^{001}-l_{i}^{100})/4
\end{equation}
where $\xi_{i}$ is a SOC parameter, while $m_{i}$ and $l_{i}$ are
atomic spin and orbital magnetic moments correspondingly.  The sum of
$\Delta_{i}$ can be compared with the total MAE $K$ obtained using the
total energies. This approach takes into account the SOC anisotropy
and its renormalization by crystal field effects. We further assume
that the spin moment has very weak anisotropy\cite{slonczewski.pr1958}
and the main change in $\xi{\bf L}\cdot {\bf S} $ product comes from
the change of orbital magnetic moment (see also
Ref.\ \onlinecite{a2}). This is the case for {\fen} (see
Table~\ref{tbl:orbmom}). For the pure {\fen}, when the spin
quantization axis rotates from $[100]$ to $[001]$, {\morb} decreases
on {${8h}$} sites, but increases on {${4d}$} and {${4e}$}. While
{\morb} depends on site, the total {\morb} increases during this
rotation, which agrees with the predicted uniaxial character of
MAE. When the spin quantization axis points along $[110]$, SOC lowers
the symmetry, and splits the four equivalent {${8h}$} sites into two
pairs with {\morb} increasing on one pair and decreasing on the other.

As shown in Fig.~\ref{fig:kmorb}, there is a strong correlation
between $K$ and {\deso} (with respect to magnetic field along
{$[001]$} direction, and atomic value $\xi_{i}$=50$m$eV is used for
all three different Fe sites for simplicity), where $i$ indicates all
atomic sites. Obviously, the atomic ${8h}$ sites make negative
contributions to the desired uniaxial MAE, and while $4e$ and $4d$
sites make positive contributions. Thus, one may hope that doping on
${8h}$ site, thus eliminating negative (in-plane) contribution to MAE,
may improve the uniaxial MAE. 

Since Co and Ti doping had been reported to stabilize the {\fen}
phase\cite{jiang.jap1999,wang.jpd1997}, it seems logical to study
prediction above using these dopants. We replaced one out of eight Fe
atoms in the primitive cell with Co or Ti atom and relax the atomic
positions within LDA and then study the anisotropy. If we replace one
of four {${8h}$} atoms with Co atom, we found that Co atom has a
larger {\morb} than any other Fe sites, however, it does not eliminate
the negative contribution from {$8h$} sites. Instead, it makes {$K$}
smaller. Also {\morb} and then {$K$} along $[100]$ and $[110]$
directions become more anisotropic. Surprisingly, however, with a Co
atom on {\fee} or {\feh} sites, the {\morb} difference between
out-of-plane and in-plane cases become even larger on {\fee} and
{\fed} sites and smaller on {\feh} sites. In other words, it makes the
positive contribution from {\fee} and {\fed} sites stronger and the
negative contribution from {\feh} sites smaller. As a result,
calculated MAE is doubled ( {$K$}=206${\times }10^{5}$erg/cm$^{3}$)
within LDA with doped Co being on {\fee} site.  A similar effect had
been found with Ti doping.  $K$ increases when Ti is substituted on
the {\fee} and {\fed} sites and decreases when substituted on {\feh}.
Unlike the Co doping, magnetic orbital moments of Ti atom are small
and barely change with spin rotation.  Generally, for the same
structure, {$K$} is always strongly correlated with {\deso}. The
larger {\deso} is, the larger {$K$} is along that specific spin
quantization direction. However, this correlation may not longer hold
true with different structures. For example, in Fe$_7$TiN, {\deso} is
the largest with Ti doped on {\feh} site, however {$K$} is much
smaller than those with Ti doped on {\fee} and {\fed} sites.

\begin{figure}[tbp]
\begin{tabular}{c}
\includegraphics[width=.48\textwidth,clip,angle=0]{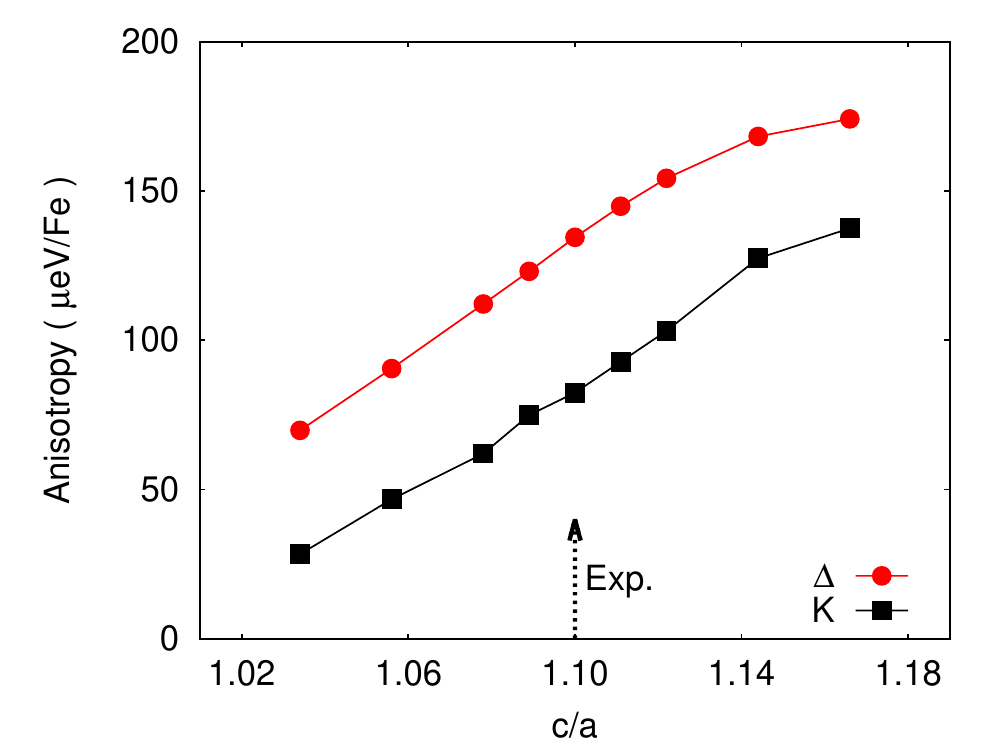}
\end{tabular}
\caption{ (Color online) $K$ and the {\deso} as functions of $c/a$ in
  {\fen}. The ideal crystal structure without strain has
  $c/a{=}1.1$. For each $c/a$, the atomic positions are relaxed with
  volume being conserved. }
\label{fig:kcoa}
\end{figure}

Tetragonality is another factor which may affect the anisotropy in a
significant way. Let us compare {\fen} with bct-Fe, where even for
{$c/a{=}1.1$} (the $c/a$ ratio for {\fen}) MAE is still rather tiny
\cite{burkert.prb2004}.  In \rfig{fig:kcoa} the calculated MAE in
     {\fen} is shown as a function of {$c/a$}. This dependence is much
     stronger than in bct-Fe and we assume that MAE mostly originates
     from distortion of Fe sublattice around the N atom and the Fe-N
     hybridization. Experimentally, the large tetragonality can be
     obtained in films, where it can be tuned by the nitrogen
     concentration\cite{ji.prb2011}. However, according to our results
     above, doping bulk {\fen} in a way that increases $c/a$ may lead
     to MAE increase. The MAE and AMAE are well correlated as shown in
     \rfig{fig:kcoa}. Within this {$c/a$} range, the spin
     magnetization varies within $ 2\%$: it is not likely to be
     responsible for the MAE increase.  On the other hand the
     anisotropy of orbital moment strongly correlates with MAE and is
     probably responsible for its enhancement as tetragonality
     increases. Orbital magnetic moments can be measured more
     precisely, so new XMCD type of experiments for this system are
     desirable.


\section{Conclusion}

In this study of intrinsic magnetic properties of {\fen}, our LDA
results for magnetization agree with previously reported values while
{\qsgw} increases magnetization by $~9\%$. This enhancement is largely
due to on-site exchange splitting between the $d$ minority and
majority states -- an effect seen in many other magnetic systems such
as NiO and MnAs \cite{chantis.prb2007,kotani.jpcm2008}. In {\fen} in
particular, we find no evidence of localized states or correlations
not already found in Fe. Taken together all of those factors we expect
that the {\qsgw} prediction for {\mave} is not far from what should be
observed in the ideal {\fen} compound. We find no evidence of charge
transfer between different Fe sites as proposed elsewhere. Thus, the
theoretical magnetization predicted for {\fen} does not exceed the
maximum on Slater-Pauling curve ($\sim$2.5\mub) and is smaller than
corresponding maximum of magnetization observed in Fe-Co alloys, which
may still be considered as a record holder among {$d$} atomic magnets.

LDA calculations predict {\tc} significantly larger than the
experimental value; the {\qsgw} result is even larger. We assume that
{\fen} will have a higher {\tc} if one can find a way to stabilize
it. Effects of doping by various elements on $M$ and {\tc} were
studied in the LMTO-CPA approximation. Various dopants affect $M$ and
{\tc} differently; but unfortunately no dopants we considered enhanced
{\mave} or {\tc}.

A uniaxial magnetocrystalline anisotropy
$K$=103$\times $$10^{5}$erg/cm$^{3}$ was calculated in the LDA with
the theoretically optimized crystal structure. $K$ is strongly
correlated with the atomic magnetic anisotropy energy due to spin-orbit
coupling only.  We found it can be increased by increasing {$c/a$} or
by adding small amount of Co or Ti atoms on {\fee} or {\fed} sites.

{\fen} is one of the more promising candidates for permanent magnets
that do not contain rare-earth elements. We believe that there is room
for improvement and we studied several possible routes to obtain
better properties. A further investigation on increasing the thermal
stability and/or changing crystal structure tetragonality is desired.

\section{Acknowledgments}

This work was supported by the U.S. Department of Energy, Office of
Energy Efficiency and Renewable Energy (EERE), under its Vehicle
Technologies Program, through the Ames Laboratory. Ames Laboratory is
operated by Iowa State University under contract
DE-AC02-07CH11358. K.\ D.\ B.\ acknowledges support from NSF through
grants DMR-1005642, EPS-1010674 (Nebraska EPSCoR), and DMR-0820521
(Nebraska MRSEC).


\bibliography{aaa}
\bigskip 

\end{document}